\documentclass[hyper]{JHEP} 
\usepackage{epsfig}
\newcommand\fverb{\setbox\pippobox=\hbox\bgroup\verb}
\newcommand\fverbdo{\egroup\medskip\noindent%
            \fbox{\unhbox\pippobox}\ }

\newcommand\fverbit{\egroup\item[\fbox{\unhbox\pippobox}]}
\newbox\pippobox

\title{D1-brane in $\beta$-Deformed Background}

\preprint{arXiv:0710.0148 [hep-th]}
\author{J. Kluso\v{n}\footnote{On leave from Masaryk University, Brno}\\
Dipartimento di Fisica,\\
Universita' \& I.N.F.N. Sezione di Roma 2, Tor Vergata \\
Via della Ricerca Scientifica 1, 00133  Roma,   ITALY\\
E-mail: \email{Josef.Kluson@roma2.infn.it}}

\author{Kamal L. Panigrahi\\
Department of Physics\\
Indian Institute of Technology Guwahati, 781 039, Guwahati, INDIA \\
E-mail: \email{panigrahi@iitg.ernet.in}}
\abstract{We study various configurations of rotating and wound
D1-brane in $AdS_5\times S^5$ background and in its $\beta$ deformed
version. We find giant magnon and spike solutions on the
world-volume of D1-brane in AdS$_5\times$ S$^5$ background. We also
analyse the equations of motion of D1-brane in $\beta$-deformed
background. We show that in the limit of large electric flux on
world-volume of D1-brane they reduce to the equations that describe
collection of large number of fundamental strings. We also construct
rotating and wound D1-brane solution that has two equal spins on
$S^5_\gamma$.}

\def \ba{\mathbf{a}}
\def \bai{\left(\ba^{-1}\right)}

\def\tgamma{\tilde{\gamma}}

\def\mF{\mathcal{F}}

\newcommand{\bA}{{\bf A}}

\newcommand{\tphi}{\tilde{\phi}}

\newcommand{\mL}{\mathcal{L}}

\newcommand{\mJ}{\mathcal{J}}

\newcommand{\bAi}{\left(\bA^{-1}\right)}

\begin{document}
\section{Introduction}
String theory on AdS$_5\times$ S$^5$ should be dual to ${\cal N}$
=4 supersymmetric Yang-Mills (SYM) theory in four
dimensions\cite{Maldacena:1997re}. This duality conjecture has
undergone varieties of tests at various levels. One of them is the
spectrum matching at the both sides of the duality. It turns out
to be very difficult to present the full spectrum of the theory
and then to compare it with spectrum of the anomalous dimension of
the gauge theory operators. Hence it is natural to examine various
limits of the duality conjecture. One such interesting class of
operators is that which carry large charges, such as large angular
momentum\cite{Berenstein:2002jq}. In this sector one can use the
semiclassical approximation to find the energy spectrum. In the
gauge theory one needs trace of very long operators. It further
was analysed in a beautiful paper \cite{Minahan:2002ve} that the
Hamiltonian of a Heisenberg's spin chain system is related with
that of the dilatation operator in N=4 Supersymmetric Yang-Mills
theory
\footnote{For recent reviews, see
 \cite{Minahan:2006sk,Plefka:2005bk,Tseytlin:2003ii}.}.
Since then lot of work has gone in to understand the
interplay between the integrability of the string theory on AdS,
and making connection with more than handful of gauge theory
operators.

One such low lying spin-chain system corresponds to magnon like
excitation. Hofman and Maldacena \cite{Hofman:2006xt} have been able
to match these magnon excitations to that of a class of
semiclassical rotating string state on $R\times S^2$ \footnote{For
some related papers, see \cite{Dimov:2007ey,Hayashi:2007bq,
Chen:2007vs,Roiban:2007jf,Papathanasiou:2007gd,Kluson:2007qu,
Minahan:2007gf,Kalousios:2006xy,Ryang:2006yq,
Hirano:2006ti,Roiban:2006gs,Chen:2006gq,Bobev:2006fg,
Spradlin:2006wk,Minahan:2006bd,Arutyunov:2006gs,Chen:2006gea,
McLoughlin:2006tz,Dorey:2006dq,Huang:2006vz}.}. They move around the
equator of the sphere and have large angular momentum and energy.
The giant magnon solution of correspond to operators where one of
the $SO(6)$ charge, $J$, is taken to infinity, keeping the $E-J$
fixed ($J\rightarrow \infty, \lambda = {\rm fixed}, p= {\rm
fixed}$). These spin chain giant magnon excitations satisfy a
dispersion relation of the type (in the large ' t Hooft limit
$(\lambda)$)
\begin{equation} E-J = \frac{\sqrt{\lambda}}{\pi}
\left|{\rm sin}\frac{p}{2}\right| \ ,
\end{equation}
where $p$ is the magnon momentum, which on the string theory side
correspond to a deficit angle $\phi$.

Another class of solutions with spike like configuration in the
AdS$_5 \times$ S$^5$ has also been found out
\cite{Kruczenski:2004wg,Kruczenski:2006pk} and in the gauge theory
side these correspond to single trace operators with large number of
derivatives. More recently in \cite{Ishizeki:2007we} is has been
analysed that for an infinitely wound string around $S^2$ and $S^3$
with a single spike can be obtained from a general solution of
rigidly rotating string on the sphere. More interestingly in a
certain parameter space the solutions can also be thought of as
giant magnon. In fact both the giant magnon and single spike
solutions has been obtained by taking different limits on a general
class of rotating solutions. So it seems natural that single spike
solutions fall into the same class of solutions as that of giant
magnon. The difference is that the spike solutions do not correspond
directly to that of any gauge theory operators. For these single
spike solution, like the magnon dispersion relation can be
summarized by
\begin{equation} E- T \Delta \psi = \frac{\sqrt{\lambda}}{\pi}
\bar{\theta} \ ,
\end{equation} where $ \Delta \psi$ is the difference in the angle
between two spikes.

It is very interesting knowing the results of the elementary
string wound around AdS$_5 \times$ S$^5$ has in its solution both
magnon and spike like configuration, what happens in case of
D-string ? It will enhance our understanding beyond the elementary
string configurations, and might also give generalization of these
as well. We would like to analyse this possibility in the present
paper. For our purpose we will restrict ourselves to the less
supersymmetric Lunin-Maldacena background \cite{Lunin:2005jy}.
This background has been conjectured to the Leigh-Strassler
marginal deformation of ${\cal N} =4 $ SYM. The study of
properties  of classical string in these backgrounds was performed
in many papers, for example
\cite{Bobev:2006fg,Ishizeki:2007we,Huang:2007th,Bobev:2007bm,Pirrone:2006iq,
Hamilton:2006ri,Huang:2006vz,Mosaffa:2007ty,Chu:2006ae,Huang:2005mi,
Bobev:2005ng,Ryang:2005pg,
Rashkov:2005mi,Frolov:2005iq,Bobev:2005cz,Frolov:2005dj,Frolov:2005ty,Chu:2007pb}.
Our goal is to perform similar analysis in the case of D1-brane.
First of all, the D-brane unlike the fundamental string couple to
the dilaton explicitly. So it is rather interesting to see if
similar configurations of giant magnon and single spike solutions
exists in case of D-branes. The corresponding operators in the
dual gauge theory is still unknown. However the existence of these
semiclassical states in string theory side might give a hint that
there might be similar states in a gauge theory. We will solve the
equations coming from the Dirac-Born-Infeld action on the
D1-string, and will analyse the possibility of getting the giant
magnon and single spike like solutions on its worldvolume.

The rest of the paper is organized as follows. In section two for
notational details we first write down the DBI action of a D1-string
and derive the equations of motion for the worldvolume coordinates
and write down the background fields corresponding to $\beta$
deformed AdS$_5 \times$ S$^5$ \footnote{For review of integrable
deformations, see \cite{Swanson:2007dh}.}. We further derive the
equations of motion of a rotating D1-brane in this background and
write down its equations of motion. Section three is devoted to the
study of the solutions of the equations of motion derived in section
two, and find out solutions that correspond to spike and giant
magnon is case of $S^2$ and $S^3$ embedded inside S$^5$. We show the
existence of the similar dispersion relation, as in case of magnon
solutions on fundamental strings, in the presence of worldvolume
gauge field. In section four we discuss the possibility of finding
out the magnon and single spike solution in case of D1-brane in
$\beta$ deformed AdS$_5 \times$ S$^5$ background.
we find more general solutions and show that in the limit of large
electric flux on world-volume of D1-brane they reduce to the
equations that describe collection of large number of fundamental
strings. We also construct rotating and wound D1-brane solution
that has two equal momenta on $S^5_\gamma$. In section five we
present our conclusions. Finally, some details of the calculations
are summarized in the appendices.
\section{D1-brane in $\beta$-Deformed Background}
The dynamics  of D1-brane in
general background is governed by following
action
\begin{eqnarray}\label{actD1}
S&=&S_{DBI}+S_{WZ} \ , \nonumber \\
S_{DBI}&=&-\tau_1
\int d^2\xi e^{-\Phi}
\sqrt{-\det\bA} \ , \nonumber \\
\bA_{\alpha\beta}&=&\partial_\alpha x^M\partial_\beta x^N
G_{MN}+(2\pi\alpha')\mF_{\alpha\beta}  \ , \nonumber \\
  \mF_{\alpha\beta}&=&
\partial_\alpha A_\beta-\partial_\beta A_\alpha-
(2\pi\alpha')^{-1}B_{MN}\partial_\alpha x^M\partial_\beta x^N
 \ , \nonumber \\
S_{WZ}&=&\tau_1\int e^{(2\pi\alpha')\mF}\wedge C \ ,  \nonumber \\
\end{eqnarray}
where $\tau_1$ is D1-brane tension, $\xi^\alpha,\alpha=0,1$ are
world-volume coordinates and where $A_\alpha$ is gauge field
living on the world-volume of D1-brane. Note also that $C$ in the
last line in (\ref{actD1}) means collection of Ramond-Ramond
fields. The equations of motion derived from this action has been
summarized in appendix-A.

Our goal is to study dynamics of D1-brane in $\beta$-deformed
$AdS_5\times S^5$ background \cite{Lunin:2005jy}. Let us now
review its main properties.

The $\beta$-deformed $AdS_5\times S^5$ background can be obtained
from pure $AdS_5\times S^5$ by a series of TsT transformations as
was shown in \cite{Frolov:2005dj}. The deformation parameter
$\beta=\gamma+i\sigma_d$ is in general a complex number however we
restrict to the case when $\sigma_d=0$, where the corresponding
deformation is called $\gamma$ deformation. The resulting
supergravity background takes the form
\begin{equation}\label{backg}
ds^2=R^2(ds^2_{AdS_5}+\sum_{i=1}^3
(d\mu_i^2+G\mu_i^2 d\phi_i^2)+\tilde{\gamma}^2 G\mu_1^2
\mu_2^2\mu_3^2(\sum_{i=1}^3 d\phi_i^2)) \ .
\end{equation}
It is important to note that this background also contains in
addition a non-trivial dilaton field as well as RR and NS-NS form
fields:
\begin{eqnarray}\label{backNS}
B&=&R^2\tilde{\gamma}G(\mu_1^2\mu_2^2d\phi_1 d\phi_2+
\mu^2_2\mu^2_3 d\phi_2 d\phi_3+\mu_1^2\mu_3^2d\phi_1 d\phi_3)
 \ , \nonumber \\
e^{2\Phi}&=&e^{2\Phi_0}G \ , \quad
G=\frac{1}{1+\tgamma^2(\mu_1^2\mu_2^2+\mu_2^2\mu_3^2+\mu_1^2\mu_3^2)} \ ,
\nonumber \\
C_2&=&-R^2\tgamma e^{-\Phi_0}\omega_1 d\psi \ , \quad
d\omega_1=12\cos\theta\sin^3\theta \sin\psi\cos\psi d\theta \wedge d\psi \ ,
\nonumber \\
F_5&=&dC_4=4R^4 e^{-\Phi_0}(\omega_{AdS_5}+\omega_{S^5}) \ ,
\nonumber \\
\mu_1&=&\sin\theta\cos\psi \ , \quad \mu_2=\cos\theta \ , \quad
\mu_3=\sin\theta\sin\psi \ , \nonumber \\
\end{eqnarray}
where $(\theta,\psi,\phi_1,\phi_2,\phi_3)$ are the usual $S^5$ variables
and where $\omega_{AdS_5},\omega_{S^5}$ are corresponding
volume forms of $AdS_5$ and $S^5$ respectively.
Finally $\tgamma$ is defined as
\begin{equation}
\tgamma=R^2\gamma \ , \quad
R^2=\sqrt{4\pi\alpha'^2 e^{\Phi_0} N} \ .
\end{equation}
Our goal is to study spikes solutions on D1-brane that moves on
$S^3_\gamma$ \footnote{various supersymmetric D-brane embeddings
in the beta deformed background was studied in
\cite{Mariotti:2007ym}. However our aim is different in this
paper}. We represent this space as a subspace of $\gamma$-deformed
$AdS_5\times S^5$ presented above
\begin{equation}
\mu_3=0 \ , \quad  \phi_3=0
\end{equation}
or equivalently
\begin{equation}
\psi=0 \ , \quad  \phi_3=0 \ .
\end{equation}
The relevant part of the $\gamma$ deformed
$AdS_5\times S^5$ is
\begin{equation}
ds^2=R^2(-dt^2+d\theta^2
+G\sin^2\theta d\phi_1^2+G\cos^2\theta d\phi_2^2)
\end{equation}
and the dilaton, RR and NS-NS two-forms take
the form
\begin{eqnarray}
B_{\phi_1\phi_2}&=&R^2\tgamma
G\sin^2\theta\cos^2\theta \ , \quad
e^{2\Phi}=e^{2\Phi_0}G \ ,  \nonumber \\
G&=&\frac{1}{1+\tgamma^2\sin^2\theta\cos^2\theta} \ .
\nonumber \\
\end{eqnarray}
Note that due to the fact that $C_2$ vanishes we do not need to
worry about the Wess-Zumino term of D1-brane effective action.

Let us now  consider
following ansatz
\begin{equation}\label{anst}
t=\kappa \tau \ , \quad \theta=\theta(y) \ ,
\quad \phi_1=\omega_1\tau+\tphi_1(y) \ , \quad
\phi_2=\omega_2\tau+\tphi_2(y) \ ,
\end{equation}
where we have defined the variable $y$ as
\begin{equation}
y\equiv \alpha \sigma+\beta\tau \ .
\end{equation}
Now we start to analyze the equations of motion (\ref{eqxm}) and
(\ref{eqa}) with the ansatz above. The explicit form of the matrix
{\bf A} is summarized in (A.6) - (A.12).

Let us see the equations of motion (\ref{eqxm}). Firstly, for
$x^0\equiv t$ this equation
 implies an existence of conserved quantity
\begin{eqnarray}\label{detbAA}
-\frac{e^{-\Phi_0}R^2\alpha^2} {\sqrt{G}\sqrt{-\det\bA}}
(G\sin^2\theta \omega_1\tphi'_1+ G\cos^2\theta\omega_2\tphi'_2)=A
\ , A=\mathrm{const} \ .
\nonumber \\
\end{eqnarray}

 In what follows we presume that $A<0$.
Further, the equations of motion for
$\phi_1$ takes the form
\begin{eqnarray}\label{eqphi1f}
-A\frac{R^2\sin^2\theta}{G}
\frac{[\omega_2
G\cos^2\theta (\tphi'_1\omega_2-\tphi'_2\omega_1)
-\phi'_1\kappa^2]}
{\sin^2\theta \omega_1\tphi'_1+\cos^2\theta
\omega_2\tphi'_2}
+\sin^2\theta\cos^2\theta\omega_2R^2\tgamma
G\Pi=B \ ,
\nonumber \\
\end{eqnarray}
where $B=\mathrm{const}$.
$\Pi$ is a constant that counts the number of fundamental strings
stretched along the world volume of D1-brane defined in (A.9).

In the same way the equation of motion for $\phi_2$ gives
\begin{eqnarray}\label{eqphi2f}
-\frac{AR^2
\cos^2\theta}
{G}
\frac{[\omega_1 G\sin^2\theta
( \tphi'_2\omega_1-\tphi'_1\omega_2)
-\tphi'_2\kappa^2]}
{\sin^2\theta \omega_1\tphi'_1+\cos^2\theta
\omega_2\tphi'_2}
-\sin^2\theta\cos^2\theta \omega_1 R^2\tgamma
G\Pi=C  \ , \nonumber \\
\end{eqnarray}
where again $C$ is constant.
Before we proceed to explicit solutions of
these equations we  determine  conserved
quantities that  reflect
  isometries of given background.
These currents  are conserved as a consequence of the
equation of motion:
\begin{equation}
\partial_\alpha \mJ^\alpha_{t,1,2}=0 \ .
\end{equation}
Then  corresponding conserved charges take the form
\begin{equation}\label{cchg}
P_t=\int_0^{2\pi}d\sigma \mJ^\tau_t \ , \quad
J_1=\int_0^{2\pi}d\sigma \mJ^\tau_1 \ , \quad
J_2=\int_0^{2\pi}d\sigma \mJ^\tau_2 \ ,
\end{equation}
where we presume that D1-brane has compact support and also that
world-volume  fields obey periodic boundary conditions.

After the general discussion of the properties of D1-brane in
$\beta$-deformed background we proceed to the study of  the
solutions of the corresponding equations of motion.
\section{D1-brane in $AdS_5\times S^5$}
We begin our discussion with the study of the dynamics of D1-brane
in original $AdS_5\times S^5$ background ($\tgamma=0$).
 For simplicity we start with
D1-brane that rotates on $S^2$. We closely follow recent
interesting paper \cite{Ishizeki:2007we}.
\subsection{Single D1-brane on $S^2$}
This case is characterised by condition
\begin{equation}
\tphi'_2=0 \ , \quad \omega_2=0 \ .
\end{equation}
Then the equation of motion for $\tphi'_1$
(\ref{eqphi1f}) implies
following  relation
\begin{equation}
R^2\kappa^2=\frac{B\omega_1}{A} \ .
\end{equation}
Since the equation (\ref{eqphi1f})
does not determine $\tphi_1$
we can presume that
\begin{equation}\label{tphi1a}
\tphi'_1=1 \
\end{equation}
so that (\ref{detbAPio}) reduces into
\begin{eqnarray}
\det\bA=-\alpha^2R^4
\frac{[\kappa^2(\theta'^2+
 \sin^2\theta)
-\theta'^2\sin^2\theta\omega_1^2]}
{1+e^{2\Phi_0}\Pi^2} \ .  \nonumber \\
\end{eqnarray}
This result together with
(\ref{detbAA})
implies following differential equation for $\theta$
\begin{eqnarray}\label{difthetas2}
 \theta'=
 \frac{\sin\theta}{|A|}
 \left[\frac{e^{-2\Phi_0}\alpha^2\omega_1^2
 (1+e^{2\Phi_0}\Pi^2)\sin^2\theta-A^2\kappa^2}
 {\kappa^2-\omega_1^2\sin^2\theta}\right]^{1/2} \ .
 \nonumber \\
\end{eqnarray}
This equation  is generalisation of
the equation  derived in paper
\cite{Ishizeki:2007we}. In fact, due to the
fact that the dilaton is constant and $C_2$
vanishes for $\gamma=0$
 the dynamics
of D1-brane has similar form as the
dynamics of fundamental string. More precisely,
it is well known that $\Pi$ determines the
number of fundamental strings on the world-volume
of D1-brane. Then the equation above determines
the dynamics of  the bound state of single
D1-brane and $\Pi$ fundamental strings  in
$AdS_5\times S^5$ background.

To proceed further  we have to impose
the boundary condition on the world-volume fields
since we  consider closed D1-brane.
The natural  boundary   conditions take the form
\begin{equation}
2\pi=\int_0^{2\pi}d\sigma=
2n\int_{\theta_{min}}^{\theta_{max}}
\frac{d\theta}{\alpha|\theta'|} \ ,
\end{equation}
where $n$ denotes  number of spikes on
D1-brane world-volume and where $\theta_{min}$ and
$\theta_{max}$ will be defined below.

Let us now evaluate charges given in
(\ref{cchg}). Using (\ref{detbAA})
we obtain that $P_t$ is equal to
\begin{eqnarray}\label{Pts2}
P_t = -\frac{2n\tau_1\kappa  R^2 }{\alpha\omega_1}
\int_{\theta_{min}}^{\theta_{max}} d\theta
\frac{\sin\theta(e^{-2\Phi_0}\alpha^2\omega_1^2
(1+e^{2\Phi_0}\Pi^2)-A^2\omega_1^2)}
{\left[(e^{-2\Phi_0}\alpha^2\omega_1^2(1+e^{2\Phi_0}
\Pi^2)\sin^2\theta-A^2\kappa^2)
(\kappa^2-\omega_1^2\sin^2\theta)\right]^{1/2}} \ .
\nonumber \\
\end{eqnarray}
In the same way we obtain
\begin{eqnarray}\label{P1s2}
P_1 =\frac{2n\tau_1 R^2}{\alpha}
\int_{\theta_{min}}^{\theta_{max}} d\theta \sin\theta
\sqrt{\frac{e^{-2\Phi_0}\alpha^2\omega_1^2
 (1+e^{2\Phi_0}\Pi^2)\sin^2\theta-A^2\kappa^2}
 {\kappa^2-\omega_1^2\sin^2\theta}} \ .
\end{eqnarray}
Finally, the difference between  two spikes
is given by an expression
\begin{equation} \label{triangles2}
\triangle \psi=
2\int_{\theta_{min}}^{\theta_{max}}
\frac{d\theta}{\alpha|\theta'|}=
-2\int_{\theta_{min}}^{\theta_{max}}
 d\theta \frac{A}{\alpha\sin\theta}
\sqrt{
\frac{\kappa^2-\omega_1^2\sin^2\theta}
{e^{-2\Phi_0}\alpha^2\omega_1^2
 (1+e^{2\Phi_0}\Pi^2)\sin^2\theta-A^2\kappa^2}} \ .
\end{equation}
Note that this is positive since we
have $A<0$.
Now requiring that the arguments in
$\theta'$ is positive we find the range
of $\theta$ can be
\begin{equation}
\mathrm{CASE \ I:} \quad
 \frac{A^2\kappa^2}{\omega_1^2
 e^{-2\Phi_0}\alpha^2
(1+e^{2\Phi_0}\Pi^2)}
< \sin^2\theta
 <\frac{\kappa^2}{\omega_1^2}
\end{equation}
or
\begin{equation}
\mathrm{CASE \ II:} \quad
\frac{\kappa^2}{\omega_1^2}
< \sin^2\theta<
 \frac{A^2\kappa^2}{\omega_1^2
 e^{-2\Phi_0}\alpha^2
(1+e^{2\Phi_0}\Pi^2)} \ .
\end{equation}
Further, in the first case we
can have $(\mathrm{i}) \frac{\kappa^2}{\omega_1^2}<1$
or $(\mathrm{ii}) \frac{\kappa^2}{\omega_1^2}>1$.
In the second case we  have
$ (\mathrm{iii}) \quad \frac{A^2\kappa^2}{
 \omega_1^2 e^{-2\Phi_0}\alpha^2
(1+e^{2\Phi_0}\Pi^2)}<1$ or
$ (\mathrm{iv}) \frac{A^2\kappa^2}{\omega_1^2
 e^{-2\Phi_0}\alpha^2
(1+e^{2\Phi_0}\Pi^2)}>1$. Note that these
results can be considered as
generalisation of  results derived
in \cite{Ishizeki:2007we}.
\subsubsection{First limiting case: giant magnon}
Let us consider  the case $(\mathrm{i})$ and $
(\mathrm{ii})$ given above and take the limit
$|\omega_1|\rightarrow \kappa$.
Following \cite{Ishizeki:2007we}
we define two angles
\begin{equation}
\sin^2\theta_{min}=
\left(\frac{A^2\kappa^2}{
\omega_1^2 e^{-2\Phi_0}\alpha^2
(1+e^{2\Phi_0}\Pi^2)}\right) \ ,
\quad
\theta_{max}=
\mathrm{arcsin}
\frac{\kappa}{\omega_1} \  , \
\theta_{min}\leq \theta \leq \theta_{max} \ .
\end{equation}
The limit $|\omega_1|\rightarrow \kappa$
corresponds to $\theta_{max}\rightarrow \frac{\pi}{2}$.
In this case the equation of motion for $\theta$
(\ref{difthetas2})
implies
\begin{equation}
\int \frac{d\theta \sin\theta_{min}
\cos\theta}{
\sin\theta\sqrt{\sin^2\theta-\sin^2\theta_{min}}}
=\pm d\sigma
\end{equation}
that has solution
\begin{equation}
\sin \theta=\mp \frac{\sin\theta_{min}}
{\sin \sigma} \ .
\end{equation}
Further, $\triangle \psi$ is equal to
\begin{eqnarray}
\triangle \psi=
2\sin\theta_{min}
\int_{\theta_{min}}^{\theta_{max}}
\frac{d\theta \cos\theta}
{\alpha\sin\theta\sqrt{\sin^2\theta-\sin^2\theta_{min}}}=
\frac{2}{\alpha}\mathrm{arcos}
(\sin\theta_{min}) \
\nonumber \\
\end{eqnarray}
and the  energy $E$
is equal to
\begin{eqnarray}
E=-P_t
=-\frac{2n\tau_1 R^2A\kappa^2}{\alpha\omega_1^2}
\int_{\theta_{min}}^{\theta_{max}}
\frac{d\theta \sin\theta(1-\sin^2\theta_{min})}{
\sin\theta_{min}
\sqrt{(\sin^2\theta-\sin^2\theta_{min})
(\sin^2\theta_{max}-\sin^2\theta)}} \ .
\nonumber \\
\end{eqnarray}
In the same way we obtain
\begin{eqnarray}
P_1=-\frac{2n\tau_1 R^2A\kappa\omega_1}{\alpha\omega^2_1}
\int_{\theta_{min}}^{\theta_{max}}
\frac{d\theta \sin\theta}{\sin\theta_{min}}
\frac{\sqrt{\sin^2\theta-\sin^2\theta_{min}}}
{\sqrt{\sin^2\theta_{max}-\sin^2\theta}}
\nonumber \\
\end{eqnarray}
and hence
\begin{eqnarray}
E-P_1=
2n e^{-\Phi_0}\tau_1\sqrt{1+e^{2\Phi_0}
\Pi^2}\sin \frac{\triangle \psi}{2} \ .
\nonumber \\
\end{eqnarray}
We have derived an analogue of the giant magnon
dispersion relation for D1-brane with world-volume electric
flux $\Pi$. Then we can interpret the solution above
as the giant magnon on the world-volume of the bound
state of single D1-brane and $|\Pi|$ fundamental strings.
\subsubsection{Second Limiting case:
Spike Solution}
The spike configuration corresponds to the
limit
\begin{equation}\label{spikelimit}
\omega^2\rightarrow
\frac{e^{-2\Phi_0} \alpha^2(1+e^{2\Phi_0}\Pi^2)}
{A^2\kappa^2}  \ .
\end{equation}
We again define
\begin{equation}
\sin\theta_{min}=\frac{\kappa}{\omega_1} \ ,
\quad
\sin^2\theta_{max}=\frac
{A^2\kappa^2}
{e^{-2\Phi_0}\omega_1^2 \alpha^2(1+e^{2\Phi_0}\Pi^2)} \
\end{equation}
so that the limit (\ref{spikelimit})
corresponds to $\theta_{max}\rightarrow \frac{\pi}{2}$.
Then the differential equation for $\theta'$
(\ref{difthetas2}) takes the form
\begin{eqnarray}
\theta'=\frac{\sin\theta_{min}\sin\theta}{\sin\theta_{max}}
\sqrt{\frac{\sin^2\theta_{max}-\sin^2\theta}
{\sin^2\theta-\sin^2\theta_{min}}} \nonumber \\
\end{eqnarray}
that can be easily integrated with the result
\begin{equation}
\frac{\cos\theta_{min}}
{\sin\theta_{min}}
\cosh^{-1}\left(\frac{\cos\theta_{min}}
{\cos\sigma}\right)
-\cos^{-1}\left(\frac{\sin\theta_{min}}{\sin\theta}
\right)
=
\pm\sigma \ .
\end{equation}
Further, for the limit (\ref{spikelimit})
the charge $P_1$  given in (\ref{P1s2})
is equal to
\begin{eqnarray}
P_1=
2n\tau_1 R^2 e^{-\Phi_0}
\sqrt{1+e^{2\Phi_0}\Pi^2}\cos\theta_{min} \ .
\nonumber \\
\end{eqnarray}
On the other hand we obtain that $E$ and $\triangle \psi$
derived in (\ref{Pts2}) and (\ref{triangles2})
diverge. However we can find
combinations of these charges that is finite
\begin{eqnarray}
E-n e^{-\Phi_0} \sqrt{1+e^{2\Phi_0}\Pi^2}\tau_1R^2 \alpha\triangle
\psi
= 2n\tau_1 R^2 e^{-\Phi_0}\sqrt{1+e^{2\Phi_0}\Pi^2}
 (\frac{\pi}{2}-\theta_{min}) \ .
\nonumber \\
\end{eqnarray}
Again, this result can be considered
as a  generalisation of the spike
relation derived in \cite{Ishizeki:2007we} to the
case of bound state of single D1-brane and collection
of $\Pi$ fundamental strings.
\subsection{D1-brane on $S^3$: two angular momenta}
Now consider more general situation when
we examine the motion of D1-brane with
an extra angular momentum. In this case the equations
of motion for $\tphi_{1}$ (\ref{eqphi1f})
and for $\tphi_{2}$  (\ref{eqphi2f})  are equal to
\begin{eqnarray}
&-&AR^2\sin^2\theta
\frac{[\omega_2
\cos^2\theta (\tphi'_1\omega_2-\tphi'_2\omega_1)
-\phi'_1\kappa^2]}
{\sin^2\theta \omega_1\tphi'_1+\cos^2\theta
\omega_2\tphi'_2}=B \ ,
\nonumber \\
&-&AR^2
\cos^2\theta
\frac{[\omega_1\sin^2\theta
( \tphi'_2\omega_1-\tphi'_1\omega_2)
-\tphi'_2\kappa^2]}
{\sin^2\theta \omega_1\tphi'_1+\cos^2\theta
\omega_2\tphi'_2}
=C \ . \nonumber \\
\end{eqnarray}
It turns out that if we combine these two equations
we obtain relations between constants $A,B$ and $C$.
In other words we are free to presume
particular form of either $\tphi_1$ or $\tphi_2$
and we choose the simplest one
 \begin{equation}
 \tphi'_1=1 \ .
 \end{equation}
Then with the help of the
 equation of motion for $\phi_1$
we find
 \begin{equation}\label{dtphi2s3}
\tphi'_2=
\frac{\sin^2\theta
(AR^2\kappa^2-B\omega_1
-AR^2\omega_2^2\cos^2\theta)}{\omega_2
\cos^2\theta(B-AR^2
\omega_1\sin^2\theta)} \ .
\end{equation}
Following  \cite{Ishizeki:2007we}
we  choose the constants of motion
appropriately so that
\begin{equation}
\theta'\rightarrow 0 \ \mathrm{as} \
\theta\rightarrow \frac{\pi}{2} \ .
\end{equation}
It turns out that the natural choice is
\begin{equation}
A=\frac{e^{-\Phi_0}\alpha\omega_1}{\kappa}
\sqrt{1+e^{2\Phi_0}\Pi^2}
 \ , \quad  B=\alpha e^{-\Phi_0}
R^2\kappa\sqrt{1+e^{2\Phi_0}\Pi^2} \ .
\end{equation}
Then the equation of motion (\ref{dtphi2s3})
simplifies considerably
\begin{equation}\label{dtphi2s3s}
\tphi'_2=\frac{\sin^2\theta \omega_1\omega_2}
{\omega_1^2\sin^2\theta-\kappa^2}  \ .
\end{equation}
Further, using (\ref{detbAA})
and (\ref{dtphi2s3s}) we easily find
differential equation for $\theta'$
\begin{eqnarray}
\theta'=\frac{|\kappa|\sin\theta\cos\theta}
{\omega_1^2\sin^2\theta-\kappa^2}
\sqrt{(\omega^2_1-\omega_2^2)\sin^2\theta-\kappa^2}
\nonumber \\
\end{eqnarray}
and consequently
\begin{eqnarray}
P_t&=&-2n\tau_1e^{-\Phi_0} R^2
\sqrt{1+e^{2\Phi_0}\Pi^2}\int_{\theta_{min}}^{\theta_{max}}
d\theta\frac{\sin \theta (\kappa^2-\omega_1^2)}
{\kappa\cos\theta \sqrt{(\omega_1^2-\omega_2^2)\sin^2\theta-\kappa^2}}
\ , \nonumber \\
P_1
&=&2n \tau_1e^{-\Phi_0} R^2\sqrt{1+e^{2\Phi_0}
\Pi^2}\int_{\theta_{min}}^{\theta_{max}}
d\theta \frac{\omega_1
\sin\theta\cos\theta}
{\sqrt{(\omega_1^2-\omega^2_2)\sin^2\theta-\kappa^2}} \ ,
\nonumber \\
P_2
&=&-2n\tau_1 e^{-\Phi_0}R^2 \sqrt{1+e^{2\Phi_0}
\Pi^2}
\int_{\theta_{min}}^{\theta_{max}}
d\theta \frac{\omega_2\sin\theta\cos\theta}
{\sqrt{(\omega_1^2-\omega^2_2)\sin^2\theta-\kappa^2}} \ .
\nonumber \\
\end{eqnarray}
Finally we find that the  difference
in angle between  two endpoints
of the string is equal to
\begin{eqnarray}
\triangle \psi
=-2\int_{\theta_{min}}^{\theta_{max}}
d\theta\frac{\omega_1^2\sin^2\theta-\kappa^2} {\alpha\kappa
\sin\theta\cos\theta\sqrt{(\omega_1^2-\omega_2^2)
\sin^2\theta-\kappa^2}}  \ .
\nonumber \\
\end{eqnarray}
In all these calculation $\theta_{max}=\frac{\pi}{2}$
and $\theta_{min}=\mathrm{arcsin}\left(\frac{|\kappa|}
{\sqrt{\omega_1^2-\omega_2^2}}\right)$ where we presume
$\omega_1^2>\omega_2^2$. Here we have chosen
$\theta_{min}$ such that insider square root is positive.
Then, since $\mathrm{arcsin}\left(\frac{|\kappa|}{
\omega_1}\right)<
\mathrm{arcsin}\left(\frac{|\kappa|}
{\sqrt{\omega_1^2-\omega_2^2}}\right)
<\pi/2, \  \theta$ can never reach a value such that
$\sin\theta=\frac{|\kappa|}{\omega_1}$. Thus in this
case $\theta'$ cannot go to infinity at any point.
Performing integrals we obtain
\begin{eqnarray}
P_1
&=&2n\tau_1 e^{-\Phi_0} R^2 \sqrt{1+e^{2\Phi_0}\Pi^2}
\frac{1}{\cos\gamma}\sin\overline{\theta} \ ,
\nonumber \\
P_2
&=&-2n\tau_1 e^{-\Phi_0} R^2 \sqrt{1+e^{2\Phi_0}\Pi^2}
\frac{\sin\gamma}{\cos\gamma}\sin\overline{\theta} \ ,
\nonumber \\
\end{eqnarray}
where we have
\begin{eqnarray}
\sin\overline{\theta}=
\frac{\sqrt{\omega_1^2-\omega_2^2-\kappa^2}}
{\sqrt{\omega_1^2-\omega_2^2}} \ ,
\quad
\sin\gamma=\frac{\omega_2}{\omega_1} \ .
\nonumber \\
\end{eqnarray}
Finally we find
\begin{eqnarray}
E-n\tau_1 R^2e^{-\Phi_0} \sqrt{1+e^{2\Phi_0}\Pi^2}\alpha\triangle
\psi
=2n\tau_1 R^2 e^{-\Phi_0} \sqrt{1+e^{2\Phi_0}\Pi^2}
\overline{\theta} \ , \nonumber \\
\end{eqnarray}
where
\begin{equation}
\overline{\theta}=
\frac{\pi}{2}-\theta_0 \ ,  \quad
\sin\theta_0=\frac{|\kappa|}
{\sqrt{\omega_1^2-\omega_2^2}} \ .
\end{equation}
Then we can also write
\begin{equation}
P_1=\sqrt{P_2^2+
2n\tau_1 R^2 e^{-\Phi_0}
\sqrt{1+e^{2\Phi_0}\Pi^2}
\sin^2\overline{\theta}} \ .
\end{equation}
Again this result can be thought of as a generalisation of the
results presented in \cite{Ishizeki:2007we} to the case of bound
state of single D1-brane and $\Pi$ fundamental strings.

In the rest of the paper we will study the dynamics of D1-brane in
$\beta$-deformed background. Before we come to this problem we
review some properties of Nambu-Goto form of the string action in
this background in appendix-B.
\section{D1-brane in $\beta$-deformed background}
In this section we return to the study of non-trivial solutions on
the world-volume of D1-brane in $\beta$-deformed background. We
closely follow the study of the fundamental string performed in
previous section. Recall that the equation of motion for $A_\alpha$
implies an existence of conserved quantity $\Pi$ defined in
(\ref{defPi}) that has the physical meaning as the number of
fundamental strings.
In analogy with the discussion performed in previous section we fix
the diffeomorphism invariance by imposing the conditions
\begin{equation}\label{d1fix}
\sqrt{-\det\bA}\sqrt{1+e^{2\Phi_0}\Pi^2G}=\bA_{\sigma\sigma} \ ,
\quad \bA_{\tau\tau}=0 \
\end{equation}
or equivalently
\begin{equation}\label{d1fixa}
(\bA_{\tau\sigma})^S=
\bA_{\sigma\sigma} \ .
\end{equation}
For this ansatz the conserved charges $P_t,J_1$ and $J_2$ and the
equations of motion for $\phi_1$ and $\phi_2$ are summarized in
appendix-C.

Using the condition  $\bA_{\tau\tau}=0$ together with the
equations of motion for $\phi_1$, $\phi_2$ in (\ref{eqphi1dg}),
(\ref{eqphi2dg}) respectively, and also the relation
(\ref{relkappa}) we obtain differential equation for $\theta'^2$
in the form
\begin{eqnarray}\label{eqthetad1ga}
\theta'^2&=&\kappa^2\frac{\beta^2+2\alpha^2-2\alpha\beta}
{(2\alpha\beta-\beta)^2}-
\nonumber \\
&-&\frac{ B^2e^{2\Phi_0}}
{R^4\sin^2\theta (1+e^{2\Phi_0}\Pi^2G)(2\alpha\beta-\beta^2)^2}
-\frac{ C^2e^{2\Phi_0}}
{R^4\cos^2\theta (1+e^{2\Phi_0}\Pi^2G)(2\alpha\beta-\beta^2)^2}-
\nonumber \\
&-&\frac{2\tgamma \alpha G e^{2\Phi_0}\Pi}
{(1+e^{2\Phi_0}\Pi^2 G)R^2(2\alpha\beta-\beta^2)^2}
(\omega_1C\sin^2\theta-\omega_2B\cos^2\theta)-
\nonumber \\
&-&\frac{\alpha^2 G(\omega_1^2\sin^2\theta+
\omega_2^2\cos^2\theta)}{(2\alpha\beta-\beta^2)^2}
\left(\frac{1+e^{2\Phi_0}\Pi^2}{1+e^{2\Phi_0}
\Pi^2G}\right) \ .
\nonumber \\
\end{eqnarray}
We must however stress one important point. The equations of
motion given above are valid in case when $ \Pi\gg 1$. This
follows from the analysis of the  equation of motion for $x^0=t$
\begin{eqnarray}\label{eqx0d}
\kappa R^2\left[\alpha \frac{e^{-\Phi}\bA_{\sigma\sigma}} {
\sqrt{-\det\bA}}- \alpha
\frac{e^{-\Phi}(\bA_{\tau\sigma})^{S}}{\sqrt{-\det\bA}}\right]'
=\kappa R^2[(\alpha-\beta)e^{-\Phi} \sqrt{1+e^{2\Phi}\Pi^2}]'=0
\end{eqnarray}
and we see that this equation is obeyed for general $\Pi$ in case
when
\begin{equation}
\alpha=\beta \ .
\end{equation}
On the other hand for $\Pi\gg 1$ we can write
$1+Ge^{2\Phi_0}\Pi^2\approx Ge^{2\Phi_0}\Pi^2$ and we see that
(\ref{eqx0d}) is automatically obeyed. Let us now analyse this
situation in more detail.
\subsection{$\Pi \gg 1$}
 This situation corresponds to
the bound state of large number of fundamental
strings $\Pi$ and one $D1$-brane. In this case
 it
is natural to  perform a rescaling $B=b\Pi \ ,
C=c\Pi$. Then in the limit $\Pi\gg 1$
 the equations (\ref{eqphi1dg}),
(\ref{eqphi2dg}) take the form
\begin{eqnarray}\label{eqphi12l}
\tphi'_1
=\frac{1}{
2\alpha\beta-\beta^2}
[\frac{
b}{R^2G\sin^2\theta}
-\alpha\omega_2 \tgamma\cos^2\theta
+(\beta\omega_1-\alpha\omega_1)] \ ,
\nonumber \\
\tphi'_2
=\frac{1}{
2\alpha\beta-\beta^2}
[\frac{c }{R^2G\cos^2\theta}
+\alpha\omega_1 \tgamma\sin^2\theta
+(\beta\omega_2-\alpha\omega_2)] \ .  \nonumber \\
\end{eqnarray}
Finally, the equation (\ref{eqthetad1ga}) reduces
into
\begin{eqnarray}\label{eqdpil}
\theta'^2&=&\kappa^2\frac{\beta^2+2\alpha^2-2\alpha\beta}
{(2\alpha\beta-\beta)^2}
-\frac{\alpha^2 (\omega_1^2\sin^2\theta+
\omega_2^2\cos^2\theta)}{(2\alpha\beta-\beta^2)^2} -
\nonumber \\
&-&\frac{ b^2}
{R^4G\sin^2\theta (2\alpha\beta-\beta^2)^2}
-\frac{c^2}
{R^4\cos^2\theta G(2\alpha\beta-\beta^2)^2}-
\nonumber \\
&-&\frac{2\tgamma \alpha }
{R^2(2\alpha\beta-\beta^2)^2}
(\omega_1c\sin^2\theta-\omega_2b\cos^2\theta) \ .
\nonumber \\
\end{eqnarray}
We see that the equations (\ref{eqphi12l})
and (\ref{eqdpil}) take exactly the same
form as the equations (\ref{eqphi12}),
(\ref{eqphi12a}) and (\ref{eqngtheta})
\footnote{After
appropriate identification of parameters
$\alpha$ and $\beta$ and constants $b,c$.}
 that describe the
dynamics of fundamental string. Further, in the limit
$e^{\Phi_0}\Pi \gg 1$  charges (\ref{chargespj}) reduce into
\begin{eqnarray}
P_t&=&-\tau_1 R^2|\Pi|\kappa\int_0^{2\pi} d\sigma
 \ ,  \nonumber \\
J_1&=&\tau_1R^2|\Pi|\int_0^{2\pi}d\sigma G\sin^2\theta
 [\omega_1+(\beta-\alpha)\tphi'_1
-\tgamma \alpha \cos^2\theta \tphi'_2]
 \ , \nonumber \\
J_2&=&\tau_1R^2|\Pi|\int_0^{2\pi}d\sigma G\cos^2\theta
[\omega_2+(\beta-\alpha) \tphi'_2 +\tgamma \alpha \sin^2\theta
\tphi'_1] \ .
\nonumber \\
\end{eqnarray}
that exactly reproduce the form of these charges for collection of
fundamental strings.
\subsection{General $\Pi$}
As we have seen above the only solution of the equation of motion
for $x^0$ corresponds to $\alpha=\beta$.
 In this case the
equation of motion for $\tphi_1,\tphi_2$ take the form
\begin{eqnarray}\label{tphi12g}
\tphi'_1=\frac{1}{\sqrt{1+e^{2\Phi_0}G \Pi^2} \alpha^2}
 [\frac{Be^{\Phi_0}}{R^2\sqrt{G}\sin^2\theta} -\alpha\omega_2\Pi
e^{\Phi_0} \tgamma\sqrt{G}\cos^2\theta ] \ .
\nonumber \\
\tphi'_2=\frac{1}{\sqrt{1+e^{2\Phi_0}G\Pi^2} \alpha^2}
[\frac{Ce^{\Phi_0}}{R^2\sqrt{G}\cos^2\theta} +\alpha\omega_1\Pi
e^{\Phi_0} \tgamma\sqrt{G}\sin^2\theta ] \ .
\nonumber \\
\end{eqnarray}
Further, the condition (\ref{d1fixa})
 for $\beta=\alpha$ implies
\begin{equation}\label{BCe}
 B\omega_1+C\omega_2=0 \ .
\end{equation}
Then, using the condition
 $ \bA_{\tau\tau}=0$ and
(\ref{BCe}) we finally obtain differential equation for $\theta$
\begin{eqnarray}\label{thetag}
\theta'^2&=&
\frac{\kappa^2}{\alpha^2}+
(\omega^2_1\sin^2\theta+\omega_2^2\cos^2\theta) \left[ \frac{2\tgamma G
e^{2\Phi_0}\Pi B} {\omega_2(1+e^{2\Phi_0}\Pi^2 G)R^2\alpha^3} \right.
-\nonumber \\
&-& \left.\frac{ B^2e^{2\Phi_0}}
{R^4\alpha^4\omega_2^2\sin^2\theta\cos^2\theta
(1+e^{2\Phi_0}\Pi^2G)}-\frac{G}{\alpha^2}
\left(\frac{1+e^{2\Phi_0}\Pi^2}{1+e^{2\Phi_0} \Pi^2G}\right)\right] \ .
\nonumber \\
\end{eqnarray}
Note also that for $\alpha=\beta$ the charges (\ref{chargespj})
take the form
\begin{eqnarray}\label{chargespjab}
P_t&=& -\tau_1 R^2 \kappa\int_0^{2\pi} d\sigma \frac{e^{-\Phi_0}}{\sqrt{G}}
\sqrt{1+e^{2\Phi_0}G\Pi^2}
 \ ,  \nonumber \\
J_1&=& \tau_1R^2\int_0^{2\pi}d\sigma[e^{-\Phi_0}\omega_1\sqrt{G}
\sin^2\theta\sqrt{1+e^{2\Phi_0} G\Pi^2}- \tgamma
G\cos^2\theta\sin^2\theta \alpha
\tphi'_2 \Pi] \ , \nonumber \\
J_2&=& \tau_1R^2\int_0^{2\pi}d\sigma[e^{-\Phi_0}\omega_2
\sqrt{G}\cos^2\theta\sqrt{1+e^{2\Phi_0} G\Pi^2}+\tgamma
G\cos^2\theta\sin^2\theta \alpha \tphi'_1 \Pi] \ .
\nonumber \\
\end{eqnarray}
It is still difficult to solve the equations (\ref{tphi12g}) and
(\ref{thetag}) for any value of $\Pi$. In fact we were not able to
find time-dependent configuration that has an interpretation as
giant magnon. For that reason we now restrict to the
case of  constant $\theta_c$.
 First of all we obtain that
$\phi_1 $ and $\phi_2$ have following solutions
\begin{eqnarray}
\phi_1&=&\omega_1\tau+\tphi'_1(\theta_c)(\alpha\sigma+\beta\tau) \
,
\nonumber \\
\phi_2&=&\omega_2\tau+\tphi'_2(\theta_c)(\alpha\sigma+\beta\tau) \
,
\nonumber \\
\end{eqnarray}
where $\tphi'_{1,2}(\theta_c)$ are constants whose explicit values
are given in (\ref{tphi12g})  evaluated for
 $\theta_c$. Note that the
periodicity conditions for $\phi_1,\phi_2$ imply
\begin{eqnarray}
\phi_1(2\pi)-\phi_1(0)&=&\tphi'_1(\theta_c)\alpha 2\pi= n_1 2\pi \
,
\nonumber \\
\phi_2(2\pi)-\phi_2(0)&=&\tphi'_2(\theta_c)\alpha 2\pi=n_2 2\pi \
,
\nonumber \\
\end{eqnarray}
where $n_{1,2}$ are winding numbers.
Further, the equation $\bA_{\tau\tau}=0$ implies the relation
between $\kappa,\omega_{1,2},n_{1,2}$ and $\theta_c$ in the form
\begin{eqnarray}\label{kappathetaconst}
0=\kappa^2(1+\tgamma^2\sin^2\theta_c\cos^2\theta_c) -\sin^2\theta_c
(\omega_1+n_1)^2
-\cos^2\theta_c(\omega_2+n_2)^2=0 \ . \nonumber \\
\end{eqnarray}
We also  have to  demand that $\theta_c$ solves the equation of motion
for $\theta$. In fact, after some algebra we obtain following
equation
\begin{eqnarray}\label{thetacon}
& &\tgamma^2(\cos^2\theta_c-\sin^2\theta_c)G
 \left[\sin^2\theta_c\omega_1^2+\cos^2\theta_c\omega_2^2+\right.\nonumber \\
&+& \left. G\sin^2\theta_c(\omega_1^2+n_1^2)e^{2\Phi_0}\Pi^2+
 G\cos^2\theta_c(\omega_2^2+n_2^2)e^{2\Phi_0}\Pi^2\right]-
 \nonumber \\
 &-&(\omega_1^2-n_1^2-\omega_2^2+n_2^2)
 \left(1+Ge^{2\Phi_0}\Pi^2\right)=0 \ .
\nonumber \\
\end{eqnarray}
Finally, if we  impose  the condition
$\bA_{\sigma\sigma}=(\bA_{\tau\sigma})^S$  we obtain
\begin{equation}
\sin^2\theta_c n_1\omega_1+\cos^2\theta_c n_2\omega_2=0 \ .
\end{equation}
We solve this  equation
 with the  ansatz
\begin{equation}\label{ansthetacon}
\omega_1=\omega_2\equiv \omega \ ,
\quad  n_1=-n_2\equiv n \ , \quad
\theta_c=\frac{\pi}{4} \ .
\end{equation}
Then it is easy to see that (\ref{thetacon}) is obeyed for the
ansatz (\ref{ansthetacon}).  In what follows we restrict ourselves
to this particular situation. Then for the ansatz
(\ref{ansthetacon})  the equation (\ref{kappathetaconst}) implies
\begin{eqnarray}\label{kappaomegan}
\kappa^2=\frac{1}{1+\frac{\tgamma^2}{4}}(\omega^2+n^2) \ .
\end{eqnarray}
Note also that for the ansatz (\ref{ansthetacon})
 the  charges (\ref{chargespjab}) take
the form
\begin{eqnarray}
P_t&=&-\tau_1 R^2 e^{-\Phi_0}2\pi
\kappa \sqrt{1+\frac{\tgamma^2}{4}+
e^{2\Phi_0}\Pi^2} \ ,
\nonumber \\
J_1&=&J_2\equiv \frac{1}{2}J=\tau_1 R^2 e^{-\Phi_0}2\pi
\frac{1} {2(1+\frac{\tgamma^2}{4})}\left[
\omega\sqrt{1+\frac{\tgamma^2}{4}+e^{2\Phi_0}\Pi^2}
+\frac{n}{2}\tgamma
e^{\Phi_0}\Pi\right] \ . \nonumber \\
\end{eqnarray}
Finally, using (\ref{kappaomegan}) we find following relation
between $E=-P_t$ and $J,\Pi$ and $n$
\begin{eqnarray}\label{EJPI}
E^2
=J^2+\left(-2\pi n \tau_1 R^2 \Pi+\frac{\tgamma}{2}J\right)^2+
(2\pi R^2\tau_1 e^{-\Phi_0}n)^2 \ .
\nonumber \\
\end{eqnarray}
First two terms above exactly reproduce the results derived in
paper \cite{Frolov:2005ty}
\footnote{The minus sign in front of $n$ is irrelevant.}
   where the term proportional to $\Pi$
describes contribution from wrapping fundamental string. The last
term in (\ref{EJPI}) follows from the contribution of wrapped
D1-brane.

\section{Conclusions}
This paper has been devoted to the study of dynamics of D1-brane in
the $AdS_5\times S^5$ background and also in its $\beta$-deformed
version. We wanted to see how D1-brane dynamics is different from
the corresponding study of the fundamental string. In case of
$AdS_5\times S^5$ we have derived the straightforward generalisation
of the giant magnon and spike solutions that were found in case of
fundamental strings \cite{Hofman:2006xt,Ishizeki:2007we}. More
precisely, we have found giant magnon and spike configurations that
are related to the dynamics of bound state of single D1-brane and
$\Pi$ fundamental strings. We mean that this is very satisfactory
result that explicitly demonstrates similarity of the classical
description of fundamental string and D1-brane in $AdS_5\times S^5$
background.

Then we proceed to the analysis of D1-brane in $\beta$-deformed
background. Now we have found that the situation is different. In
fact, we showed that in case of the large number of fundamental
strings that are stretched along world-volume of D1-brane the
dynamics of this system takes the same form as in case of the
fundamental string \cite{Bobev:2007bm}. This result again
demonstrates the consistency of our approach. On the other hand in
case of finite number of fundamental strings we were not able to
find time dependent configurations that could be interpreted as
giant spikes or magnons on the world-volume of D1-brane. We mean
that this is a consequence of the fact that classical D1-brane
explicitly couples to dilaton which is non-trivial in the
$\beta$-deformed background and has significant contribution to the
dynamics of D1-brane. On the other hand when we have restricted to
the study of dynamics of D1-brane with constant $\theta$ we have
been able to find the generalisation of the formula derived in
\cite{Frolov:2005ty}.

In summary, we hope that our result could be useful for further
study of the dynamics of D1-brane in $AdS_5\times S^5$ background
and its deformation. It would be certainly interesting to study
properties of D1-brane in the $\beta$-deformed $AdS_5\times S^5$
background with complex deformation parameter.

\section*{Acknowledgements}

J.K. would like to thank Institute of Theoretical Physics and
Astrophysics, Masaryk University, Brno   for hospitality where
part of  this work has been done. This work  was supported in part
by the Czech Ministry of Education under Contract No. MSM
0021622409, by INFN, by the MIUR-COFIN contract 2003-023852 and ,
by the EU contracts MRTN-CT-2004-503369 and MRTN-CT-2004-512194,
by the INTAS contract 03-516346 and by the NATO grant
PST.CLG.978785.

\begin{appendix}
\section{Equations of motion}\label{A}
Varying the action (\ref{actD1}) with respect to $x^M$ we obtain
the following equations of motion for $x^M$ \footnote{equations of
motion for all the branes in AdS spacetime has been discussed in
\cite{Skenderis:2002vf}}
\begin{eqnarray}\label{eqxm}
& &-\tau_1\partial_M[ e^{-\Phi}]
\sqrt{-\det\bA}-\nonumber \\
&-&\frac{\tau_1}{2} e^{-\Phi}(\partial_M g_{KL}\partial_\alpha x^K
\partial_\beta x^L-
\partial_M b_{KL}
\partial_\alpha x^K\partial_\beta x^L)\bAi^{\beta\alpha}
\sqrt{-\det\bA}+\nonumber \\
&+&\tau_1\partial_\alpha [e^{-\Phi}g_{MN}\partial_\beta
x^N\bAi_S^{\beta\alpha}\sqrt{-\det\bA}]-
\nonumber \\
&-&\tau_1
 \partial_\alpha
[e^{-\Phi}b_{MN}\partial_\beta x^N
\bAi_A^{\beta\alpha}\sqrt{-\det\bA}]+J_M=0 \ , \nonumber \\
\end{eqnarray}
where
\begin{equation}
J_M=\frac{\delta S_{WZ}}{\delta x^M} \
\end{equation}
and where
\begin{equation}\label{bASA}
\bAi_S^{\alpha\beta}=
\frac{1}{2}\left(\bAi^{\alpha\beta}+\bAi^{\beta\alpha}\right) \ ,
\quad \bAi_A^{\alpha\beta}=
\frac{1}{2}\left(\bAi^{\alpha\beta}-\bAi^{\beta\alpha}\right) \ .
\end{equation}
In the same way the variation of (\ref{actD1}) with respect to
$A_\alpha$ implies following equation of motion
\begin{equation}\label{eqa}
2\pi\alpha'\tau_1
\partial_\beta [e^{-\Phi}\bAi^{\beta\alpha}_A\sqrt{-\det\bA}]+
J^\alpha=0 \ ,
\end{equation}
where
\begin{equation}
J^\alpha=\frac{\delta S_{WZ}}{\delta A_\alpha} \ .
\end{equation}

Now we  start to analyse the equations of motions given in
(\ref{eqxm}) and (\ref{eqa}). To begin with note that for the
ansatz (\ref{anst}) the matrix $\bA$ is equal to
\begin{eqnarray}
\bA_{\tau\tau}&=&R^2[ -\kappa^2+\beta^2\theta'^2+
G\sin^2\theta(\omega_1+\beta\tphi'_1)^2+
G\cos^2\theta(\omega_2+\beta\tphi'_2)^2] \ ,
\nonumber \\
\bA_{\tau\sigma}&=& R^2[\alpha\beta \theta'^2+ G\sin^2\theta
\alpha (\omega_1+\beta \tphi'_1)\tphi'_1+ G\cos^2\theta
\alpha(\omega_2+\beta\tphi'_2)\tphi'_2+
\nonumber \\
&+& \tgamma G\sin^2\theta\cos^2\theta
\alpha(\omega_2\tphi'_1-\omega_1\tphi'_2)]+2\pi\alpha'
F \ , \nonumber \\
\bA_{\sigma\tau}&=& R^2[\alpha\beta\theta'^2+G\sin^2\theta
\alpha(\omega_1+\beta\tphi'_1)\tphi'_1+
G\cos^2\theta\alpha(\omega_2+\beta\tphi'_2)\tphi'_2-\nonumber \\
&-& \tgamma G\sin^2\theta\cos^2\theta \alpha
(\omega_2\tphi'_1-\omega_1\tphi'_2)]-2\pi\alpha' F \ ,
\nonumber \\
\bA_{\sigma\sigma}&=& R^2[\alpha^2\theta'^2+G\sin^2\theta \alpha^2
\tphi'^2_1+
G\cos^2\theta \alpha^2\tphi'^2_2] \ , \nonumber \\
\end{eqnarray}
where $F_{\tau\sigma}\equiv F$ and where $(\dots)'\equiv
\frac{d}{dy}(\dots)$. Then it is easy to calculate $\det\bA$ and
we obtain
\begin{eqnarray}
\det \bA&=& -\alpha^2R^4\kappa^2
[\theta'^2+G\sin^2\theta\tphi'^2_1+
G\cos^2\theta\tphi'^2_2]+\nonumber \\
&+& \alpha^2R^4G^2\cos^2\theta \sin^2\theta
(\omega_1\tphi'_2-\omega_2\tphi'_1)^2+
\nonumber \\
&+&\alpha^2 R^4 G\theta'^2(\sin^2\theta\omega_1^2+
\cos^2\theta\omega_2^2)+
\nonumber \\
&+& [\tgamma R^2G\sin^2\theta\cos^2\theta
\alpha(\omega_2\tphi'_1-\omega_1\tphi'_2)+2\pi\alpha'
F]^2 \ .  \nonumber \\
\end{eqnarray}
Let us now return to the equations of motion (\ref{eqxm}) and
(\ref{eqa}).
 The
equation of motion for $A_\alpha $ implies
\begin{eqnarray}
\partial_\tau[e^{-\Phi}
\bAi^{\tau \sigma}_A\sqrt{-\det\bA}]=0 \ ,
\nonumber \\
\partial_\sigma[e^{-\Phi}
\bAi^{\sigma\tau}_A\sqrt{-\det\bA}]=0
\nonumber \\
\end{eqnarray}
and consequently
\begin{equation}\label{defPi}
e^{-\Phi}\frac{(\bA_{\tau\sigma})_A}{\sqrt{-\det\bA}}= \Pi \ ,
\end{equation}
where $\Pi$ is constant that counts the number of fundamental
strings stretched along world-volume of D1-brane. Further, using
the properties of matrices $\bA^S$ and $\bA^A$ defined in
(\ref{bASA}) we easily find
\begin{eqnarray}\label{detA}
\det\bA=\bA_{\tau\tau}\bA_{\sigma\sigma}-
\bA_{\tau\sigma}\bA_{\sigma\tau}=
\nonumber \\
\bA_{\tau\tau}\bA_{\sigma\sigma}-
(\bA_{\tau\sigma})^S(\bA_{\tau\sigma})^S
+(\bA_{\tau\sigma})^A(\bA_{\tau\sigma})^A \ .
\nonumber \\
\end{eqnarray}
If we combine (\ref{defPi})  with (\ref{detA}) we can express
$(\bA_{\tau\sigma})^A$ as
\begin{eqnarray}
(\bA_{\tau\sigma})^A(\bA_{\tau\sigma})^A (e^{-2\Phi}+\Pi^2)=
(-\bA_{\tau\tau}\bA_{\sigma\sigma}+
(\bA_{\tau\sigma})^S(\bA_{\tau\sigma})^S)\Pi^2
\nonumber \\
\end{eqnarray}
and hence the determinant $\det\bA$  takes the form
\begin{eqnarray}\label{detbAPio}
\det\bA
&=&\frac{\alpha^2R^4}{ 1+Ge^{2\Phi_0}\Pi^2}\times
\left[\kappa^2(\theta'^2+
 G(\sin^2\theta\tphi'^2_1+
\cos^2\theta\tphi'^2_2))-\right.\nonumber \\
&-&\left.G^2\cos^2\theta\sin^2\theta
(\omega_1\tphi'_2-\omega_2\tphi'_1)^2
-\theta'^2G(\sin^2\theta\omega_1^2+
\cos^2\theta\omega_2^2)\right]\ . \nonumber \\
\end{eqnarray}

\subsection{Conserved Charges}
The conserved currents are given by
\begin{eqnarray}
\mJ^\alpha_t&=& \frac{\delta \mL}{\delta \partial_\alpha t}
=-\tau_1 e^{-\Phi}g_{tt}\partial_\beta t\bAi^{\beta\alpha}_S
\sqrt{-\det\bA} \ , \nonumber \\
\mJ^\alpha_1&=&\frac{\delta \mL}{\delta \partial_\alpha\phi_1}=
-\tau_1 e^{-\Phi}[ g_{\phi_1\phi_1}\partial_\beta
\phi_1\bAi^{\beta\alpha}_S + b_{\phi_1\phi_2}\partial_\beta \phi_2
\bAi^{\beta\alpha}_A]\sqrt{-\det\bA} \ , \nonumber \\
\mJ^\alpha_2&=& \frac{\delta \mL}{\delta\partial_\alpha \phi_2}=
-\tau_1 e^{-\Phi}[ g_{\phi_2\phi_2}\partial_\beta
\phi_2\bAi^{\beta\alpha}_S + b_{\phi_2\phi_1}\partial_\beta \phi_1
\bAi^{\beta\alpha}_A]\sqrt{-\det\bA} \  . \nonumber \\
\end{eqnarray}
\section{Nambu-Goto form of the string action in
$\beta$-deformed background} In order to understand better the
dynamics of D1-brane in $\beta$-deformed background we consider
Nambu-Goto action for fundamental string in this background. Our
goal is to explicitly see how analysis of this action can be
related to the analysis of the sigma model form of the action
presented in \cite{Bobev:2007bm}.

Let us start with the Nambu-Goto action for fundamental string in
general background
\begin{equation}\label{NGaction}
S=-\frac{1}{2\pi\alpha'} \int d\tau d\sigma[ \sqrt{-\det
\ba_{\alpha\beta}}+\frac{1}{2}\varepsilon^{\alpha\beta}
b_{MN}\partial_\alpha x^M\partial_\beta x^N] \ ,
\end{equation}
where $\varepsilon^{\alpha\beta}=-\varepsilon^{\beta\alpha} \ ,
\quad \varepsilon^{01}=1 \ , \quad  \ba_{\alpha\beta}=
g_{MN}\partial_\alpha x^M\partial_\beta x^N $. Variation of
(\ref{NGaction}) with respect to  $x^M$ implies following equation
of motion
\begin{eqnarray}
& &\frac{1}{2}\partial_M g_{KL}\partial_\alpha x^K
\partial_\beta x^L \bai^{\beta\alpha}\sqrt{-\det\ba}
-\partial_\alpha [g_{MN}\partial_{\beta}x^N
\bai^{\beta\alpha}\sqrt{-\det\ba}]+\nonumber \\
&+&\frac{1}{2}\varepsilon^{\alpha\beta}
\partial_M b_{KL}\partial_\alpha x^K\partial_\beta x^L-
\partial_\alpha[\varepsilon^{\alpha\beta}b_{MN}
\partial_\beta x^N]=0 \ . \nonumber \\
\end{eqnarray}
For reader's convenience we again write the  relevant part of the
$\beta$-deformed $AdS_5\times S^5$ background
\begin{equation}
ds^2=R^2(-dt^2+d\theta^2 +G\sin^2\theta d\phi_1^2+G\cos^2\theta
d\phi_2^2)
\end{equation}
and
\begin{eqnarray}
B_{\phi_1\phi_2}=R^2\tgamma G\sin^2\theta\cos^2\theta \ , \quad
e^{2\Phi}= e^{2\Phi_0}G \ , \quad
G=\frac{1}{1+\tgamma^2\sin^2\theta\cos^2\theta} \ .
\nonumber \\
\end{eqnarray}
Let us now consider following ansatz
\begin{equation}\label{ansNG}
t=\kappa \tau \ , \quad \theta=\theta(y) \ , \quad
\phi_1=\omega_1\tau+\tphi_1(y) \ , \quad
\phi_2=\omega_2\tau+\tphi_2(y) \ ,
\end{equation}
where
\begin{equation}
y\equiv \alpha \sigma+\beta\tau \ .
\end{equation}
For this ansatz components of matrix $\ba$ take the form
\begin{eqnarray}
\ba_{\tau\tau}&=&R^2[-\kappa^2+\beta^2
\theta'^2+G\sin^2\theta(\omega_1+\beta\tphi'_1)^2+ G\cos^2\theta
(\omega_2+\beta\tphi'_2)^2]
\ ,  \nonumber \\
\ba_{\tau\sigma}&=&\ba_{\sigma\tau}= R^2
[\alpha\beta\theta'^2+G\sin^2\theta
\alpha(\omega_1+\beta\tphi'_1)\tphi'_1+ G\cos^2\theta
\alpha(\omega_2+\beta\tphi')\tphi'_2] \ ,
\nonumber \\
\ba_{\sigma\sigma}&=& R^2\alpha^2[\theta'^2+G\sin^2\theta
\tphi'^2_1+
G\cos^2\theta \tphi'^2_2] \ \nonumber \\
 \end{eqnarray}
 and consequently
\begin{eqnarray}
\det \ba&=& -\alpha^2R^4\kappa^2
[\theta'^2+G\sin^2\theta\tphi'^2_1+
\nonumber \\
&+& G\cos^2\theta\tphi'^2_2]+\alpha^2R^4G^2\cos^2\theta
\sin^2\theta (\omega_1\tphi'_2-\omega_2\tphi'_1)^2+
\nonumber \\
&+&\theta'^2\alpha^2G(\sin^2\theta\omega_1^2+
\cos^2\theta\omega_2^2) \ .
\nonumber \\
\end{eqnarray}
Note that the fundamental string has following conserved currents
\begin{eqnarray}
\mJ_t^\alpha&=&\frac{\mL_{NG}}{\partial_\alpha t}=
-\frac{1}{2\pi\alpha'}g_{tt}\partial_\beta t\bai^{\beta\alpha}
\sqrt{-\det\ba} \ ,
\nonumber \\
\mJ_{\phi_1}^\alpha&=& \frac{\mL_{NG}}{\partial_\alpha \phi_1}=
-\frac{1}{2\pi\alpha'}g_{\phi_1\phi_1}
\partial_\beta \phi_1\bai^{\beta\alpha}
\sqrt{-\det\ba}-\frac{1}{2\pi\alpha'}
\varepsilon^{\alpha\beta}b_{\phi_1\phi_2}
\partial_\beta\phi_2 \ ,
\nonumber \\
\mJ_{\phi_2}^\alpha&=& \frac{\mL_{NG}}{\partial_\alpha \phi_2}=
-\frac{1}{2\pi\alpha'}g_{\phi_2\phi_2}
\partial_\beta \phi_2\bai^{\beta\alpha}
\sqrt{-\det\ba} -\frac{1}{2\pi\alpha'}\varepsilon^{\alpha\beta}
b_{\phi_2\phi_1}\partial_\beta\phi_1 \ .
\nonumber \\
\end{eqnarray}
Then for the ansatz (\ref{ansNG}) we obtain following form of
conserved charges $P_t,J_1,J_2$
\begin{eqnarray}\label{chargeNG}
P_t&=&
-\frac{R^2\kappa}{2\pi\alpha'} \int_0^{2\pi}d\sigma \frac{
\ba_{\sigma\sigma}}
{\sqrt{-\det\ba}} \ , \nonumber \\
J_1&=&
\frac{1}{2\pi\alpha'} \int_0^{2\pi}d\sigma
\frac{R^2G\sin^2\theta}{\sqrt{-\det\ba}}
[(\omega_1+\beta\tphi'_1)\ba_{\sigma\sigma}-
\alpha \tphi'_1 \ba_{\tau\sigma}]-\nonumber \\
&-&\frac{1}{2\pi\alpha'}\int_0^{2\pi} d\sigma R^2 \tgamma \alpha
G\sin^2\theta
\cos^2\theta \tphi'_2 \ ,\nonumber \\
J_2&=&
\frac{1}{2\pi\alpha'} \int_0^{2\pi}d\sigma
\frac{R^2G\cos^2\theta}{\sqrt{-\det\ba}}
[(\omega_2+\beta\tphi'_2)\ba_{\sigma\sigma}-
\alpha \tphi'_2 \ba_{\tau\sigma}]+\nonumber \\
&+&\frac{1}{2\pi\alpha'}\int_0^{2\pi} d\sigma R^2\tgamma \alpha
G\sin^2\theta
\cos^2\theta \tphi'_1 \ . \nonumber \\
\nonumber \\
\end{eqnarray}
Nambu-Goto action is still diffeomorphism invariant. We fix this
gauge freedom by demanding that
\begin{eqnarray}
\sqrt{-\det\ba}=\sqrt{-\ba_{\tau\tau}\ba_{\sigma\sigma}+
\ba_{\tau\sigma}^2}=\ba_{\sigma\sigma}
\end{eqnarray}
that can be solved with the condition \footnote{It is clear that
we could use an  alternative gauge fixing solution
$\ba_{\tau\tau}=-\ba_{\sigma\sigma} \ , \ba_{\tau\sigma}=0$.}
\begin{equation}\label{rashkovc1}
\ba_{\tau\tau}=0
\end{equation}
and hence
\begin{equation}\label{rashkovc2}
\ba_{\sigma\sigma}=\ba_{\tau\sigma} \ , \quad
\sqrt{-\det\ba}=\ba_{\tau\sigma} \ .
\end{equation}
With this gauge fixing the charges given in (\ref{chargeNG}) are
equal to
\begin{eqnarray}
P_t&=& -\frac{R^2\kappa}{2\pi\alpha'}
\int_0^{2\pi}d\sigma  \ , \nonumber \\
J_1&=& \frac{R^2}{2\pi\alpha'} \int_0^{2\pi} d\sigma
G\sin^2\theta[\omega_1-\tphi'_1\frac{a}{\kappa} -\alpha\tgamma
\cos^2\theta\tphi'_2]
\nonumber \\
J_2&=& \frac{R^2}{2\pi\alpha'} \int_0^{2\pi} d\sigma
G\cos^2\theta[\omega_2-\tphi'_2\frac{a}{\kappa} +\alpha\tgamma
\sin^2\theta\tphi'_1] \ .
\nonumber \\
\end{eqnarray}
Then  using (\ref{rashkovc1}) and (\ref{rashkovc2}) it is easy to
see that the equation of motion for $\phi_1$ implies following
differential equation for $\tphi'_1$
\begin{eqnarray}\label{eqphi12}
\tphi'_1=\frac{1}{\beta^2-2\beta\alpha}
\left[\frac{b}{R^2G\sin^2\theta}- \alpha\omega_2 \tgamma
\cos^2\theta-
\omega_1(\beta-\alpha)\right] \ ,  \nonumber \\
\end{eqnarray}
where $b$ is constant. Note that if
 we choose the parametrisation
$\beta-\alpha=-\beta'$ we obtain
\begin{equation}
\tphi'_1=\frac{1}{(\beta'^2-\alpha^2)}
[\frac{b}{R^2G\sin^2\theta}- \alpha\omega_2 \gamma \cos^2\theta+
\omega_1\beta'_1]
\end{equation}
that coincides exactly with the equations of motion given in
\cite{Bobev:2007bm}. In the same way the equation of motion for
$\phi_2$ implies
\begin{eqnarray}\label{eqphi12a}
\tphi'_2=\frac{1}{\beta^2-2\beta\alpha}
[\frac{c}{R^2G\cos^2\theta}- \omega_2(\beta-\alpha)+\alpha
\omega_1\tgamma
\sin^2\theta] \ , \nonumber \\
\end{eqnarray}
where $c$ is again constant.

In order to find differential equation for $\theta$ we use the
condition $\ba_{\tau\tau}=0$ together with (\ref{eqphi12}) and
(\ref{eqphi12a}) and we obtain
\begin{eqnarray}\label{eqngtheta}
\theta'^2
&=&\frac{1}{(\beta^2-2\beta\alpha)^2}
[\kappa^2(\beta^2-2\beta\alpha+2\alpha^2) -\frac{b^2}{R^4
G\sin^2\theta}-\frac{c^2}{R^4 G\cos^2\theta}+
\nonumber \\
&+&\frac{2\tgamma\alpha}{R^2}
(\omega_2b\cos^2\theta-\omega_1c\sin^2\theta)
-\alpha^2(\omega_1^2\sin^2\theta+\omega_2^2\cos^2\theta)] \ ,
\nonumber \\
\end{eqnarray}
where we have also used the relation
\begin{equation}\label{rel}
0=\kappa^2+\frac{ \omega_2 c}{R^2(\alpha- \beta)}+ \frac{ \omega_1
b}{R^2(\alpha-\beta)} \ .
\end{equation}
This relation follows from the condition
$\ba_{\tau\sigma}=\ba_{\sigma\sigma}$ that implies
\begin{eqnarray}
\theta'^2(\alpha\beta-\alpha^2)
+\alpha\omega_1G\sin^2\theta\tphi'_1+\alpha\omega_2
G\cos^2\theta\tphi'_2= (\alpha^2-\alpha\beta)
[G\sin^2\theta\tphi'^2_1+G\cos^2\theta\tphi'^2_2] \ .
\nonumber \\
\end{eqnarray}
Then if we combine this result with the condition
 $\ba_{\tau\tau}=0$ and use
 (\ref{eqphi12}) and (\ref{eqphi12a}) we finally
obtain (\ref{rel}).

It is easy to see that  if we make the substitution
$\beta=\alpha-\beta'$ in (\ref{eqngtheta}) we obtain the same
equation that was presented in \cite{Bobev:2007bm}. A careful
analysis presented there shows that there exist two solutions
corresponding to giant magnon and spikes. We will not repeat these
calculations here and recommend the original paper
\cite{Bobev:2007bm} for more details.

\section{Conserved charges for the D1-brane in $\beta$-
deformed background} The conserved charges $P_t, J_1, J_2$ for the
D1-brane in the $\beta$ deformed background can be calculated as
\begin{eqnarray}\label{chargespj}
P_t&=&
-\tau_1 R^2 \kappa\int_0^{2\pi} d\sigma
\frac{e^{-\Phi_0}}{\sqrt{G}} \sqrt{1+e^{2\Phi_0}G\Pi^2}
 \ ,  \nonumber \\
J_1&=&
\tau_1R^2\int_0^{2\pi}d\sigma[e^{-\Phi_0}\sqrt{G}\sin^2\theta
 (\omega_1+(\beta-\alpha)\tphi'_1)\sqrt{1+e^{2\Phi_0}
G\Pi^2}-\nonumber \\
&-& \tgamma G\cos^2\theta\sin^2\theta \alpha
\tphi'_2 \Pi] \ , \nonumber \\
J_2&=&
\tau_1R^2\int_0^{2\pi}d\sigma[e^{-\Phi_0}\sqrt{G}\cos^2\theta
 (\omega_2+(\beta-\alpha)\tphi'_2)\sqrt{1+e^{2\Phi_0}
G\Pi^2}+\nonumber \\
&+&\tgamma G\cos^2\theta\sin^2\theta \alpha \tphi'_1 \Pi] \ .
\nonumber \\
\end{eqnarray}
 Further,
the equations of motion for $\phi_1$ reduces to
\begin{eqnarray}\label{eqphi1dg}
\tphi'_1&=&\frac{1}{\sqrt{1+e^{2\Phi_0}G \Pi^2}(
2\alpha\beta-\beta^2)}\times
\nonumber \\
&\times & [\frac{Be^{\Phi_0}}{R^2\sqrt{G}\sin^2\theta}
-\alpha\omega_2\Pi e^{\Phi_0} \tgamma\sqrt{G}\cos^2\theta
+(\beta\omega_1-\alpha\omega_1) \sqrt{1+e^{2\Phi_0}G\Pi^2}] \ .
\nonumber \\
\end{eqnarray}
In the same way the equation of motion for $\phi_2$ gives
\begin{eqnarray}\label{eqphi2dg}
\tphi'_2&=&\frac{1}{\sqrt{1+e^{2\Phi_0}G\Pi^2}(
2\alpha\beta-\beta^2)}\times
\nonumber \\
&\times& [\frac{Ce^{\Phi_0}}{R^2\sqrt{G}\cos^2\theta}
+\alpha\omega_1\Pi e^{\Phi_0} \tgamma\sqrt{G}\sin^2\theta
+(\beta\omega_2-\alpha\omega_2) \sqrt{1+e^{2\Phi_0}G\Pi^2}]
\nonumber \\
\end{eqnarray}
Now if we combine the diffeomorphism invariance condition
(\ref{d1fixa}) together with the condition $\bA_{\tau\tau}=0$ we
obtain the relation
\begin{eqnarray}\label{relkappa}
0=\kappa^2-\frac{\sqrt{G}e^{\Phi_0}}{(\alpha-\beta)
\sqrt{1+e^{2\Phi_0}G\Pi^2}R^2} [B\omega_1+ C\omega_2]  \ .
\end{eqnarray}

\end{appendix}

\end{document}